\documentclass{PoS}
\usepackage{graphicx}
\usepackage{gensymb}

\title{LHC results on tree-level beauty decays}

\ShortTitle{LHC results on tree-level beauty decays}

\author{\speaker{Mika Vesterinen}\thanks{On behalf of the LHCb Collaboration.}\\
        University of Oxford\\
        E-mail: \email{mika.vesterinen@cern.ch}}

\abstract{Tree-level $b$ decays play a critical role in characterising the quark flavour sector,
and exposing possible effects of physics beyond the Standard Model.
These proceedings cover recent results from the LHCb experiment on
semileptonic $b$ baryon decays, $\mathcal{R}(D^{\ast-})$ using three-prong hadronic $\tau$ decays,
$CP$ observables in $B^- \to D^{(\ast)}h^-$ decays, and an updated combination on the CKM {angle~$\gamma$}.
}

\FullConference{The European Physical Society Conference on High Energy Physics\\
		5-12 July, 2017\\
		Venice}

\begin{document}

\section{Introduction}

Precision studies of $b$ hadron decay properties offer a great opportunity to discover indirect effects of physics beyond the Standard Model (SM).
Flavour changing neutral current processes are especially promising since the SM is only capable of mediating them
through loop-suppressed amplitudes and therefore beyond SM (BSM) amplitudes have a fighting chance of leaving a measurable imprint on observables.
While tree-level decays are generally expected to be immune to BSM effects, 
they play a critical role in exposing anomalous behaviour in loop observables.
Furthermore, tree-level decays to final states with third generation leptons are 
directly sensitive to BSM scenarios with non universal leptonic couplings.

An attractive null test of the SM is presented through the unitarity of the 
Cabibbo Kobayashi Maskawa (CKM) matrix, which relates the quark mass eigenstates to their corresponding flavour eigenstates~\cite{PhysRevLett.10.531,doi:10.1143/PTP.49.652}.
Six of the unitarity constraints are of the type $\sum_{k}V_{ik}V_{jk}^{*} = 0$, which can be represented as triangles,
the areas of which are related to the extent of $CP$ violation in the corresponding $ij$ quark sectors.
The $bd$ triangle has the largest area and is commonly referred to as {\em the} Unitarity Triangle (UT),
and a core focus of quark flavour physics is to overconstrain its apex using a combination of tree-level and loop-level observables.
The loop-level determination of the apex is currently more precise.
It is therefore of paramount importance to improve the precision of the tree-level constraints.

One side of the triangle is proportional to $|V_{ub}|/|V_{cb}|$, which can be determined using 
tree-level semileptonic decays of the type $b \to q \ell \nu_{\ell}$ ($q = c,u$).
In Sect.~\ref{sec:SLB} several recent studies of semileptonic decays of $b$ baryons, by the LHCb experiment, are discussed.
Another powerful application of tree-level $b$ decays is to confront the SM assertion of lepton universality.
Sect.~\ref{sec:RD} considers the status of these measurements and reports on a very recent analysis by LHCb,
which is the first to use three-prong hadronic $\tau$ decays.
Decays that are mediated by interfering tree-level $b \to c\bar{u}s$  and $b \to \bar{c}us$ amplitudes 
allow a theoretically clean determination of the internal UT angle $\gamma = \arg(-V_{ud}V_{ub}^*/V_{cd}V_{cb}^{*})$.
In Sect.~\ref{sec:BToDK} a new analysis of $B^- \to D^{0(\ast)}K^-$ decays is presented,
and Sect.~\ref{sec:gamma} reports on a determination of $\gamma$ from a combination of LHCb analyses.





\section{\label{sec:SLB}Studies of semileptonic beauty baryon decays by LHCb} 

Semileptonic decays of $b$ hadrons are ideally suited to extract $|V_{ub}|$ and $|V_{cb}|$ since they 
only involve one hadronic current that can be expressed through form factors which can 
be computed using lattice QCD (LQCD) methods.
Extensive studies of semileptonic decays of $B$ mesons have been performed by $e^+e^-$ collider experiments.
Determinations of $|V_{ub}|$ and $|V_{cb}|$ have been made using 
both inclusive and exclusive measurements, but there is an uncomfortable pattern of tensions between them~\cite{Amhis:2016xyh},
though recent progress has been made in understanding this matter~\cite{Grinstein:2017nlq,Bigi:2017njr}.



\begin{figure}\centering
\includegraphics[width=0.45\textwidth]{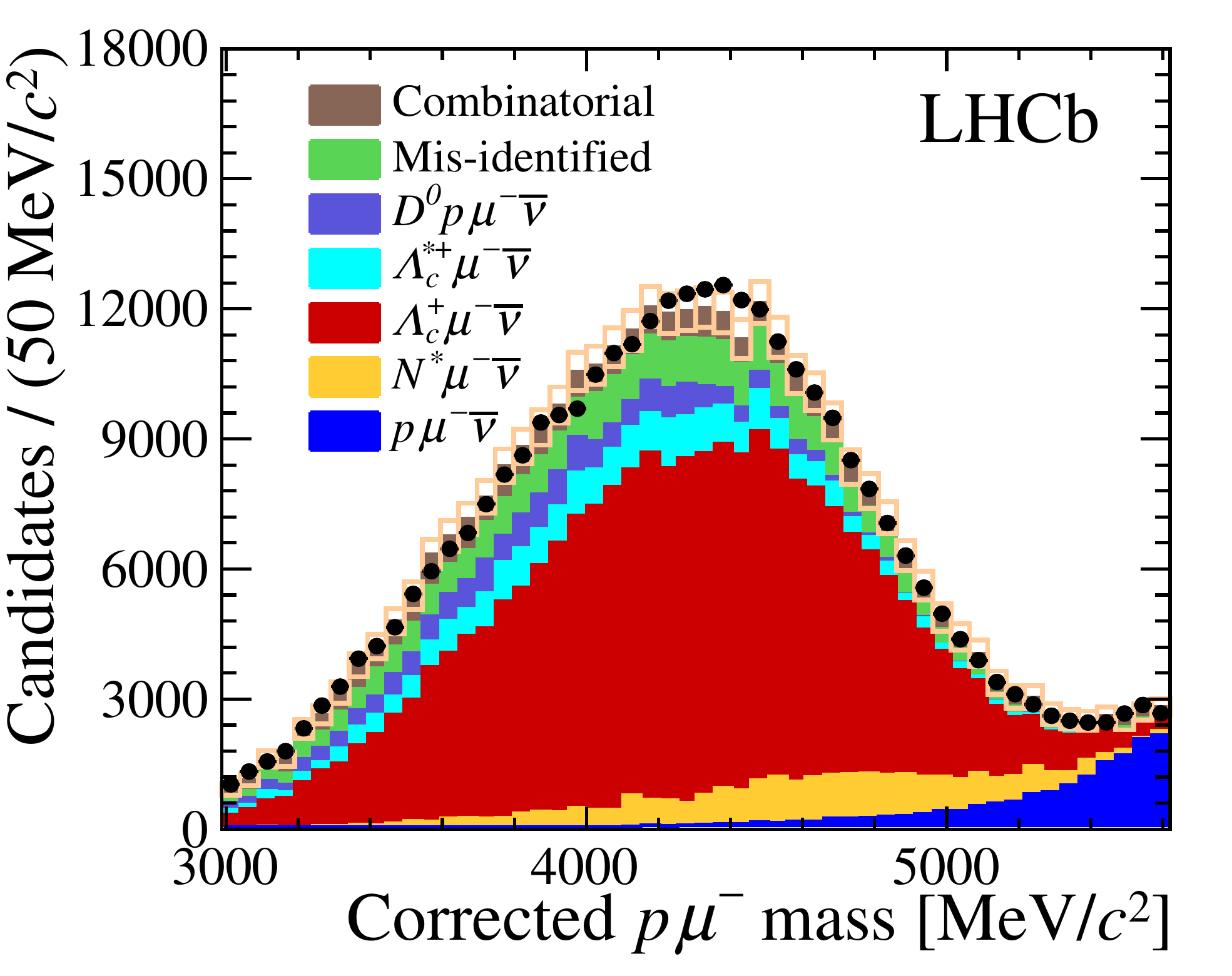}
\includegraphics[width=0.49\textwidth]{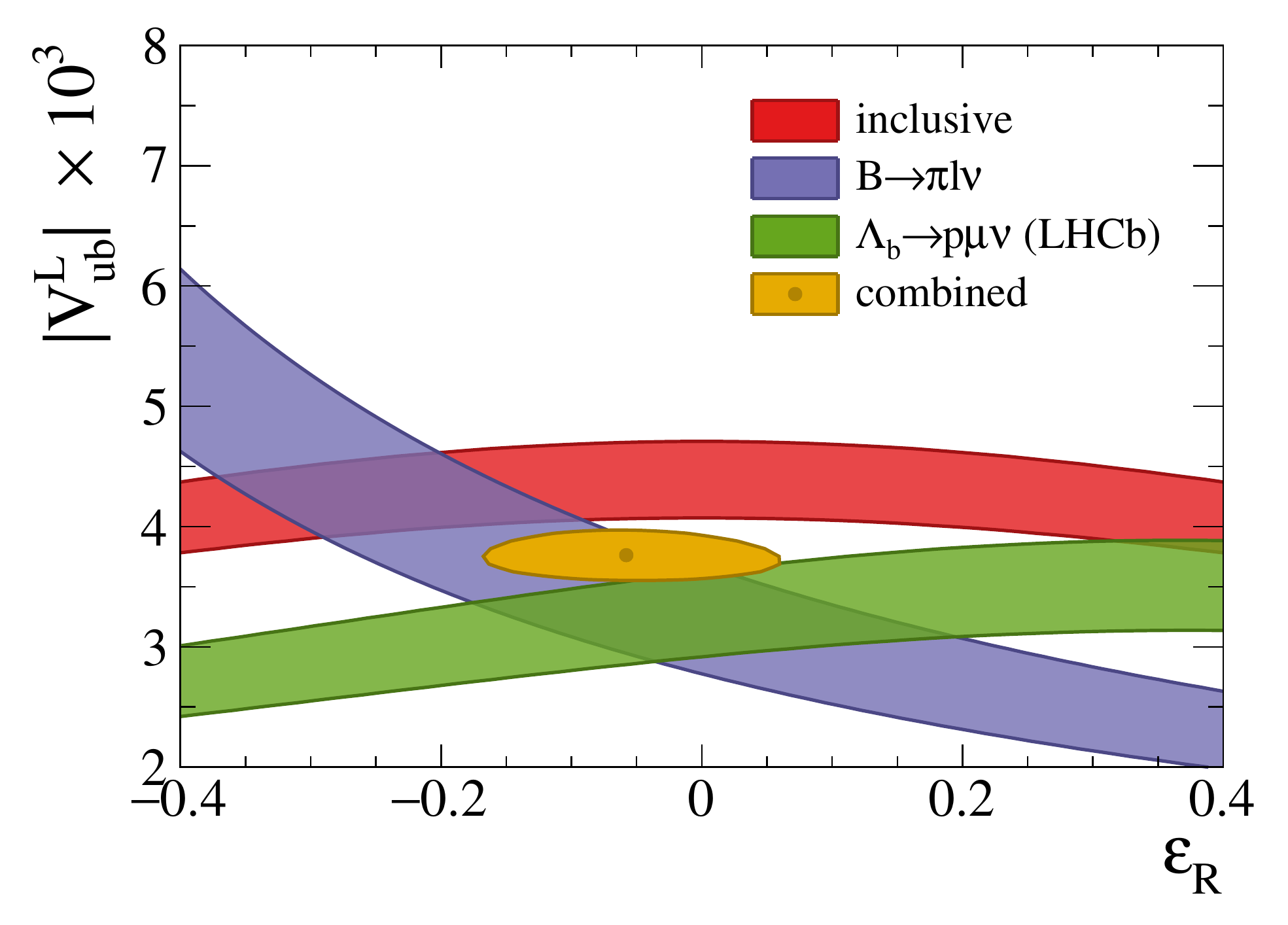} 
\caption{\label{Fig:pmu}Left: A template fit to the corrected mass distribution of candidate $\Lambda_b^0 \to p\mu^- \overline{\nu}_{\mu}$ decays.
Right: The dependence of different $|V_{ub}|$ determinations on the strength of an anomalous right-handed current.}
\end{figure}

The LHCb experiment presents a compelling opportunity to make complementary studies with higher statistics,
and importantly, with $b$ hadron species that cannot be produced at $e^+e^-$ colliders operating at the $\Upsilon(4S)$ resonance.
LHCb is particularly well suited to measure ratios of $b \to c\ell\nu_{\ell}$ versus $b \to u\ell\nu_{\ell}$ transitions,
with the same $b$ hadron species, since the poorly known $b$ cross section cancels.
Motivated by precise LQCD calculations of the associated form factors~\cite{Detmold:2015aaa},
a first such study with $\Lambda_b^0$ baryon decays was reported by LHCb in 2015, based on 2~fb$^{-1}$ of integrated luminosity~\cite{Aaij:2015bfa}.
Fig~\ref{Fig:pmu} (left) shows a template fit to the distribution of the corrected mass variable, as defined for example in Ref.~\cite{Aaij:2015bfa},
of $\Lambda_b^0 \to p\mu^- \overline{\nu}_{\mu}$ candidates.
A significant first observation of this signal is clearly evident to the right-hand side of the corrected mass distribution.
The same variable can be used to statistically separate the exclusive $\Lambda_b^0 \to \Lambda_c^+ \mu^- \overline{\nu}_{\mu}$ component,
from the various feed-down contributions, allowing a determination of the ratio,
\begin{equation}
  \frac{\mathcal{B}(\Lambda_b^0 \to p \mu^-\overline{\nu}_{\mu})_{q^2 > 15~\mathrm{GeV}^2/c^4}}{\mathcal{B}(\Lambda_b^0 \to\Lambda_c\mu^-\overline{\nu}_{\mu})_{q^2 > 7~\mathrm{GeV}^2/c^4}} = \left( 1.00 \pm 0.04_{\rm stat} \pm 0.08_{\rm syst} \right) \times 10^{-2}.
\end{equation}
Using the LQCD form factors~\cite{Detmold:2015aaa}, this ratio is translated to the following determination,
\begin{equation}
\frac{|V_{ub}|}{|V_{cb}|} = 0.083 \pm 0.004_{\rm expt} \pm {0.004}_{\rm lattice}.
\end{equation}
The precision of this determination is comparable to previous studies with $B$ mesons,
and it favours the exclusive determinations of $|V_{ub}|$ and $|V_{cb}|$.
It can be seen in Fig.~\ref{Fig:pmu} (right) how anomalous right-handed currents, as suggested in Ref.~\cite{Bernlochner:2014ova},
could relieve the inclusive--exclusive $|V_{ub}|$ tension.
Baryonic decays provide a very powerful test of this explanation, due to their complementary spin structure.
The LHCb result disfavours this scenario.


\begin{figure}\centering
\includegraphics[width=0.49\textwidth]{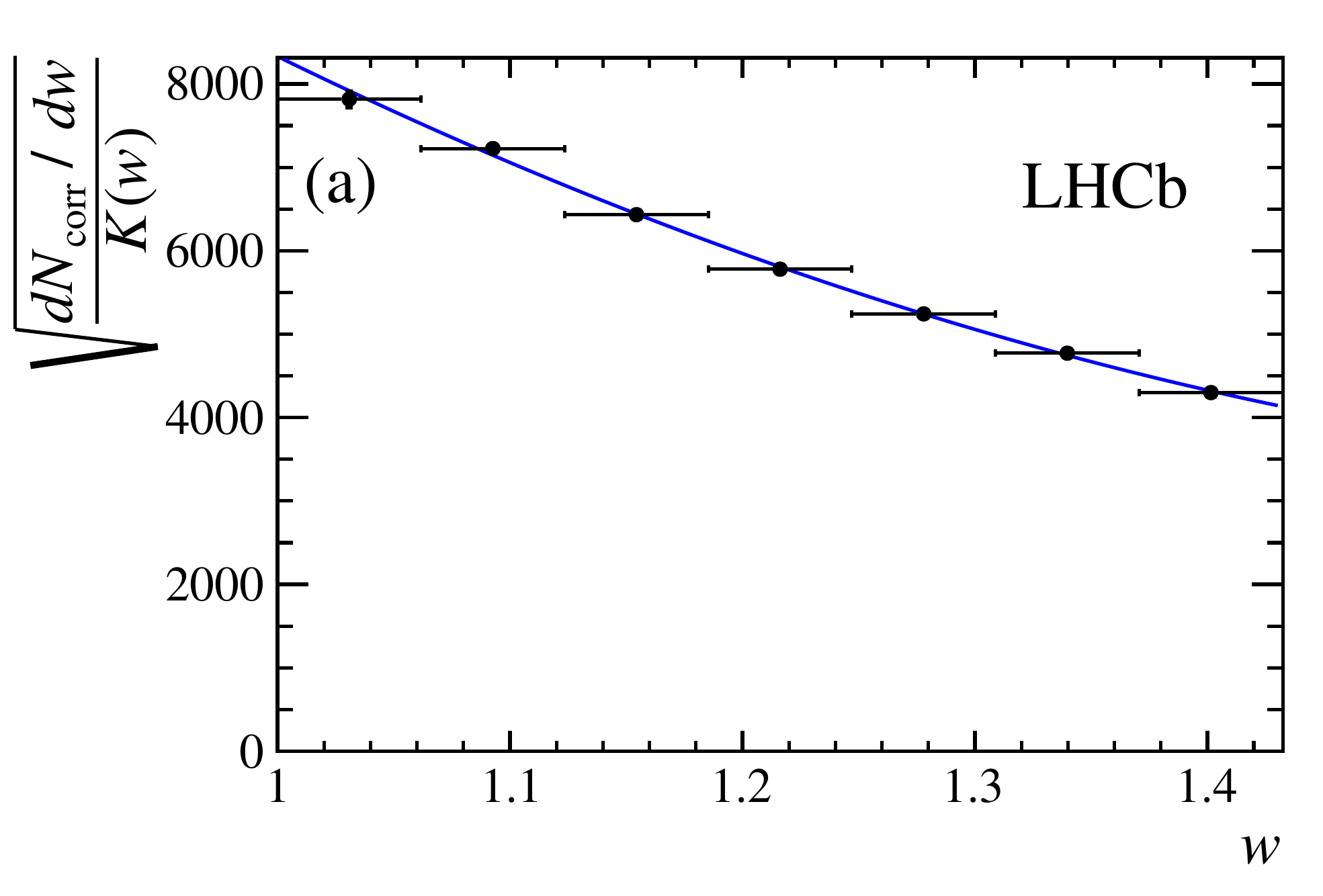}
\includegraphics[width=0.49\textwidth]{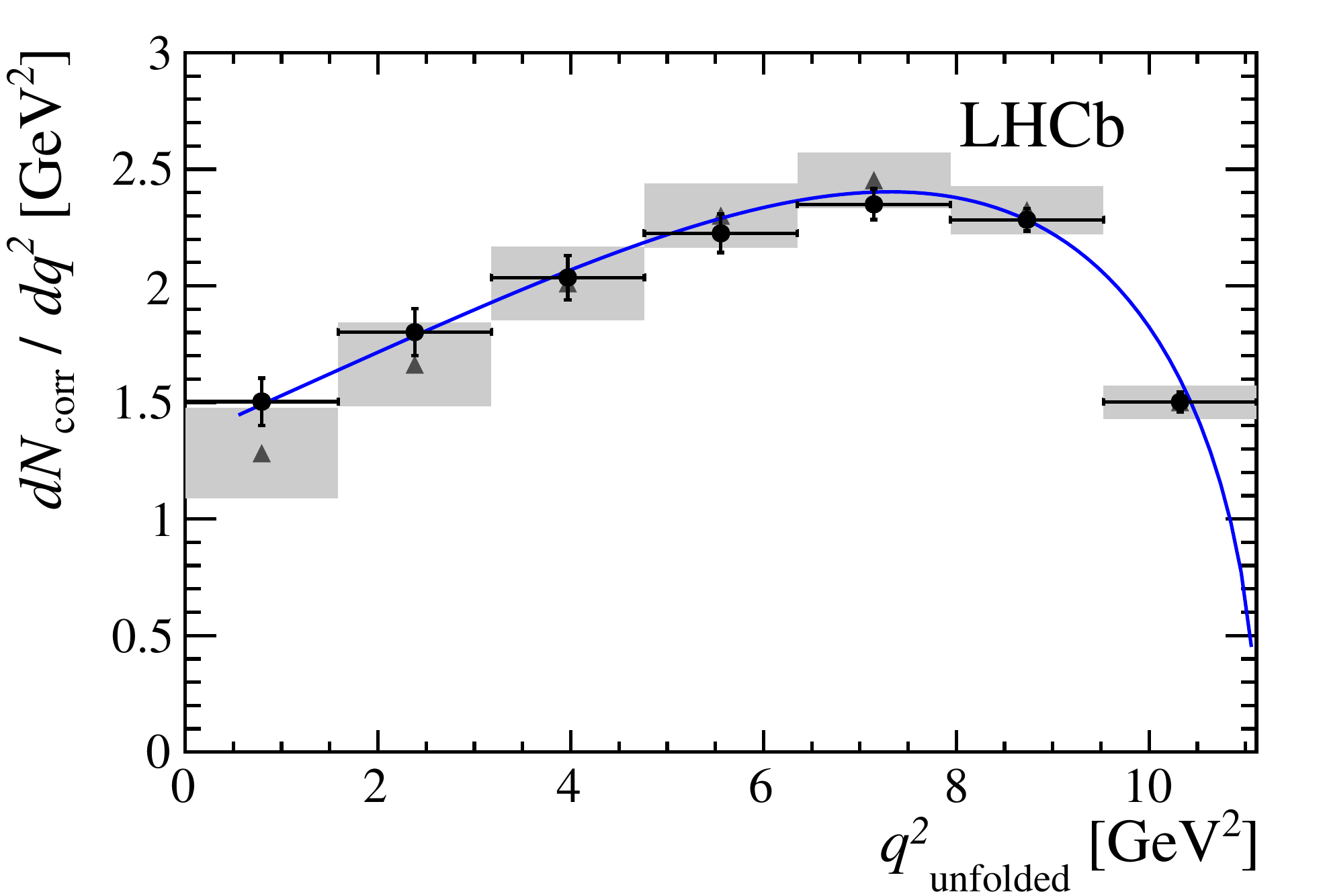}
\caption{\label{Fig:Lc}Left: The measured Isgur-Wise function of the decay $\Lambda_b^0 \to \Lambda_c^+\mu^-\overline{\nu}_{\mu}$.
Right: The unfolded $q^{2}$ distribution, as compared to LQCD predictions.}
\end{figure}

LHCb is well also suited to measure the normalised differential decay rates of semileptonic decays.
A detailed shape study of $\Lambda_b^0 \to \Lambda_c^+\mu^-\overline{\nu}_{\mu}$ decays, 
with the full Run-I dataset, is reported~\cite{Aaij:2017svr}.
In the heavy quark limit, it can be shown that all form factors describing the semileptonic decay of a heavy flavoured
hadron are proportional to a universal function known as the Isgur-Wise (IW) function~\cite{ISGUR1989113}.
The decay $\Lambda_b^0 \to \Lambda_c^+\mu^-\overline{\nu}_{\mu}$ is described by six independent form factors,
but in the static approximation of infinite quark masses, these can be expressed in terms of a single 
heavy-baryon IW function $\zeta(w)$, where $w = v_{\Lambda_b^0}\cdot v_{\Lambda_c^0}$.
Prior to LHCb, there has only been one study of semileptonic $\Lambda_b^0$ decays by the DELPHI experiment at LEP~\cite{Abdallah:2003gn}, with limited precision.
Fig.~\ref{Fig:Lc} (left) shows the measured IW function, unfolded for experimental resolution and efficiency effects.
The line indicates a fit of the IW function expanded to second order in $(w-1)$,
with the parameters $\rho^2$ and $\sigma^2$ describing the slope and curvature, respectively, 
of $\zeta(w)$ at $w=1$.
The fit reports $\rho^2 = 1.63 \pm 0.07$ and $\sigma^2 =  2.16 \pm 0.34$ with a correlation coefficient of 0.97.
This precise result supports predictions of LQCD~\cite{Bowler:1997ej}, QCD sum rules~\cite{Huang:2005mea}, and the relativistic quark model~\cite{Ebert:2006rp}.
The shape of the differential decay rate is also measured as a function of $q^2$, and is found to agree well with the LQCD
predictions~\cite{Detmold:2015aaa} that were used in the $|V_{ub}|/|V_{cb}|$ extraction, as can be seen in Fig.~\ref{Fig:Lc} (right).

Many studies of semileptonic $b$ decays are in progress or are planned.
For example a study of $B_s^0$ meson decays to $K^-\mu^+ \nu_{\mu}$ and $D_s^-\mu^+\nu_{\mu}$ final states is strongly motivated by 
the precision of the form factors of the $V_{ub}$ mode from LQCD~\cite{Flynn:2015mha}.
New ideas have also been suggested to improve the resolution of the decay kinematics,
thus permitting differential measurements with finer binning~\cite{Ciezarek:2016lqu}.

\section{\label{sec:RD}Lepton universality tests with tree-level semileptonic decays}

Ratios of $b \to c \tau \nu$ decay rates, with respect to the corresponding channels with light leptons,
for example,
\begin{equation}
\mathcal{R}(D^{(\ast)}) = \frac{\mathcal{B}(B \to D^{(\ast)} \tau \nu_{\tau})}{\mathcal{B}(B \to D^{(\ast)} \ell \nu_{\ell})},
\end{equation}
can be computed with high precision since hadronic uncertainties are cancelled to a large degree~\cite{PhysRevD.85.094025,Bigi:2016mdz}.
Meanwhile they are highly sensitive to BSM scenarios in which new forces have preferential couplings to third generation leptons.

The summer 2016 HFLAV averages of $\mathcal{R}(D)$ and $\mathcal{R}(D^*)$, combining measurements
from BaBar~\cite{Lees:2012xj,Lees:2013uzd}, Belle~\cite{Huschle:2015rga,Sato:2016svk}, and LHCb~\cite{Aaij:2015yra},
are 3.9$\sigma$ away from the SM predictions~\cite{Amhis:2016xyh}.
More recently, Belle published a first simultaneous measurement of $\mathcal{R}(D^{\ast})$ and the $\tau$ polarisation using single-prong hadronic $\tau$ decays~\cite{Hirose:2016wfn}.

The first LHCb study~\cite{Aaij:2015yra} was based on muonic $\tau$ decays.
Recently, LHCb presented a measurement of $\mathcal{R}(D^{*-})$ using the three-prong hadronic $\tau$ decay ($\tau^- \to \pi^-\pi^-\pi^+(\pi^0)\nu_{\tau}$)~\cite{Aaij:2017uff},
which is the first study using this mode.
This decay channel has several advantages over the muonic mode.
For example, there is no background from $B \to D^{\ast-}\mu^+\nu_{\mu}$ decays.
The decay topology also allows the $\tau$ decay vertex to be reconstructed.
The first challenge is to suppress the background from $B \to D^{*-}\pi^+\pi^-\pi^-X$ decays,
which swamps the signal by two orders of magnitude, prior to any dedicated attempt to suppress it.
The excellent resolution of the LHCb vertex detector allows this background to be almost entirely eliminated 
by requiring the $\tau$ decay vertex to be significantly downstream of the $B$ decay vertex.
The background from $B \to D^{*-}X_cX$ decays, with $X_c \to \pi^+\pi^-\pi^+ X$, survives this requirement,
and still requires special treatment.
A multivariate discriminant is developed to distinguish these decays
from the signal, taking advantage in particular  of the very different resonant substructure 
of the three pion system between the signal and the $D_s^+ \to \pi^+\pi^-\pi^+ X$ events which form the largest background source.
The region with the exclusive decay  $D_s^+ \to \pi^+\pi^-\pi^+$ provides a very clean sample to 
control the cocktail of feed-downs to the inclusive $B \to D^{*-}D_{s}^+X$ decay.

\begin{figure}\centering
\includegraphics[width=0.6\textwidth]{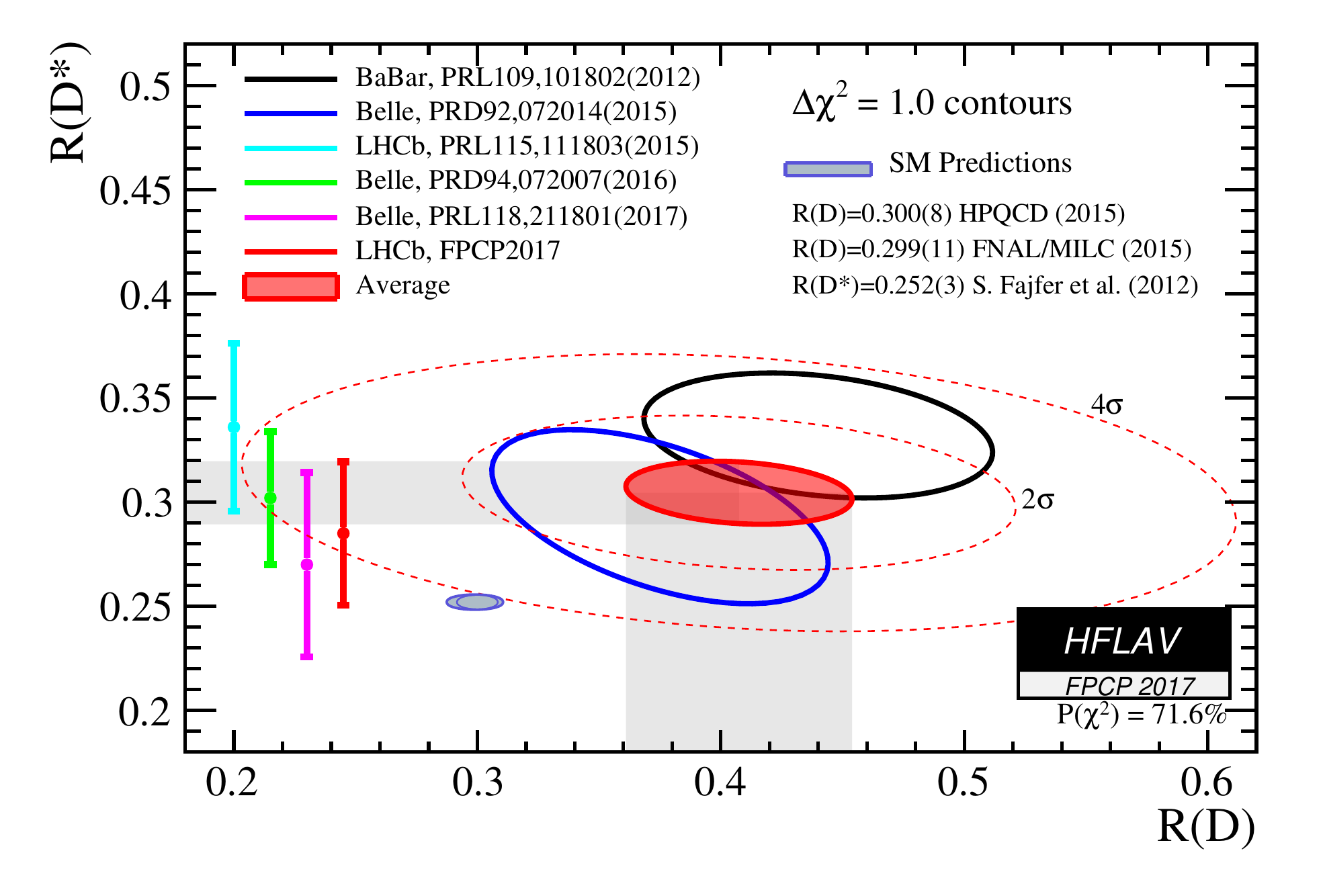} 
\caption{\label{Fig:RD}Measurements of $\mathcal{R}(D^{\ast})$ and $\mathcal{R}(D)$, and their averages,
compared to the SM predictions, as compiled by the HFLAV collaboration.}
\end{figure}

The signal yield is extracted through a fit in fine bins of  $\tau$ decay time and $q^2$, and four coarse bins in 
the multivariate discriminant output.
In the highest purity of these four bins the signal purity is roughly 50\%.
This promises excellent prospects for precise characterisation of the decay with higher statistics future LHCb datasets.
The branching ratio is measured with respect to a topologically related normalisation mode, as follows
\begin{equation}
  \frac{ \mathcal{B}(B^0 \to D^{*-}\tau^+\nu_{\tau}) }{\mathcal{B}(B^0 \to D^{*-} \pi^+\pi^+\pi^-) } = 1.93 \pm 0.13_{\rm stat} \pm 0.17_{\rm syst}.
\end{equation}
Using the measured branching ratios for the normalisation mode~\cite{1674-1137-40-10-100001},
and the well known $B^0 \to D^{*-}\mu^+\nu_{\mu}$ branching ratio~\cite{Amhis:2016xyh},
a value of 
\begin{equation}
\mathcal{R}(D^{*-}) = 0.285 \pm 0.019_{\rm stat} \pm 0.025_{\rm syst} \pm 0.013_{\rm norm},
\end{equation}
is obtained.
Fig.~\ref{Fig:RD} depicts the state-of-the-art for $\mathcal{R}(D^{(*)})$ determinations,
as compared to the precise SM predictions, and including this recent three-prong result from LHCb.
While the new LHCb result is compatible with the SM prediction for $\mathcal{R}(D^{*})$, it is also consistent with the
other experimental measurements, and the overall tension in the two-dimensional plane remains at a very similar level ($4.1$ standard deviations).
Studies are ongoing within LHCb towards a series of similar measurements with other $b$-hadron species.

\section{\label{sec:BToDK}Measurement of \boldmath{$CP$} observables with \boldmath{$B^{\pm} \to D^{0(\ast)}h^{\pm}$} decays}

The CKM angle $\gamma$ can be cleanly determined through a study of the family of 
$B \to D K$ decays.
The decay $B^- \to D^0K^-$ proceeds through a $b \to c\bar{u}s$ transition,
while the equivalent decay with a $\overline{D}^0$ proceeds through $b \to \bar{c}us$.
The interference between these two paths, which occurs when the $D^0$ or $\overline{D}^0$ decay to the same final state,
is sensitive to the phase $\gamma$.
For example, $D^0$ or $\overline{D}^0$ decays to the $CP$ eigenstates $K^+K^-$ and $\pi^+\pi^-$ are well suited~\cite{GRONAU1991172,GRONAU1991483}.
The resulting $B^-$ decays exhibit a $CP$-violating asymmetry whose magnitude depends on $\gamma$,
and on the relative amplitude and strong phase of the two paths.
LHCb has previously analysed this class of decay modes with the Run-I dataset~\cite{Aaij:2016oso}.

\begin{figure}\centering
\includegraphics[width=0.8\textwidth]{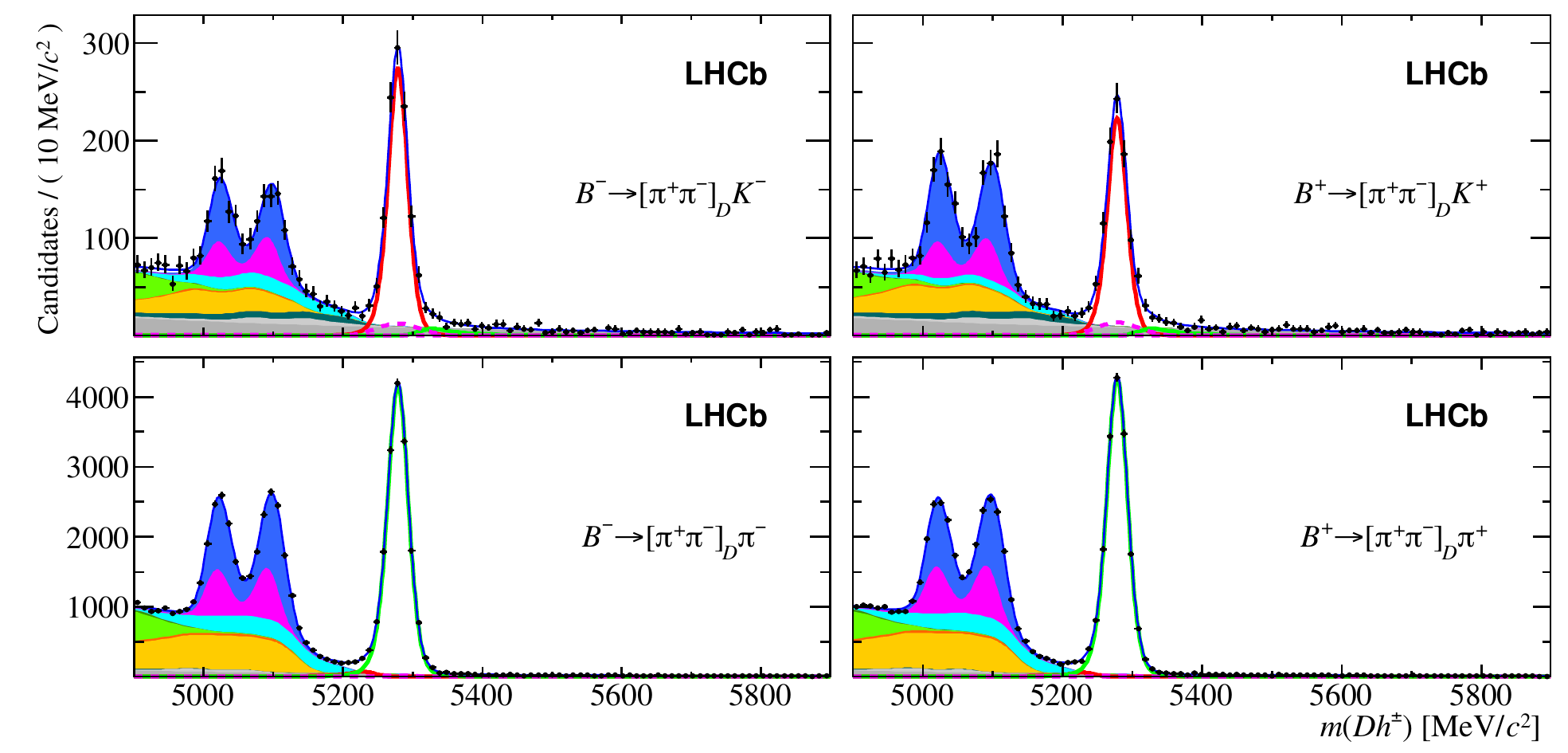} 
\includegraphics[width=0.8\textwidth]{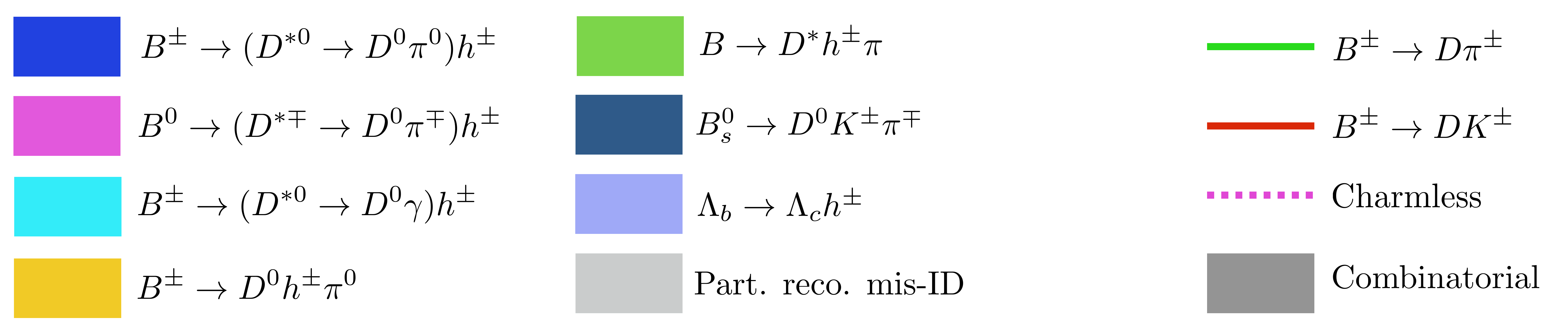}
\caption{\label{Fig:GLW}The invariant mass spectra of the selected $B^- \to Dh^-$ ($h=\pi,K$) candidates.}
\end{figure}

It has also been suggested to study the equivalent observables with $B^- \to D^{\ast 0} K^-$ decays
since there is an exact strong phase difference of $\pi$ between the decays with
$D^{\ast 0} \to D^0 \gamma$ compared to $D^{\ast 0} \to D^0\pi^0$~\cite{Bondar:2004bi}.
LHCb recently reported an update of~\cite{Aaij:2016oso} including 2~fb$^{-1}$ of Run-II data~\cite{Aaij:2017ryw},
and the measurement of $D^{\ast 0}$ observables.
Fig.~\ref{Fig:GLW} shows the invariant mass spectra of the $B^- \to D^0h^-$ ($h=\pi,K$) candidate samples, 
with  $D^0 \to \pi^+\pi^-$.
The explicit reconstruction of the $D^{\ast 0}$ decays is challenging, but it can be seen in Fig.~\ref{Fig:GLW}
that they form distinct structures below the main $B^-$ peaks, when {\em partially reconstructed} under the 
$B^- \to Dh^-$ hypotheses.
A clear $CP$ asymmetry can be seen between the signal yield of $B^- \to D^0K^-$ versus $B^+ \to D^0K^+$.
Similar asymmetries are seen in the partially reconstructed signals, but the asymmetries
of the $D^{\ast 0} \to D^0 \gamma$ mode has the opposite sign to that of the  $D^{\ast 0} \to D^0\pi^0$ mode,
which is expected given their relative strong phase shift mentioned above.
There is no clear evidence for $CP$ violation in the $B^- \to D^0\pi^-$ modes.
The determination of a total of 19 observables is reported in~\cite{Aaij:2017ryw}, 
of which eight correspond to the fully reconstructed $B^- \to D^0 h^-$ decays, and the remainder to the 
partially reconstructed decays. 

\section{\label{sec:gamma}LHCb \boldmath{$\gamma$} combination}

\begin{figure}\centering
\includegraphics[width=0.6\textwidth]{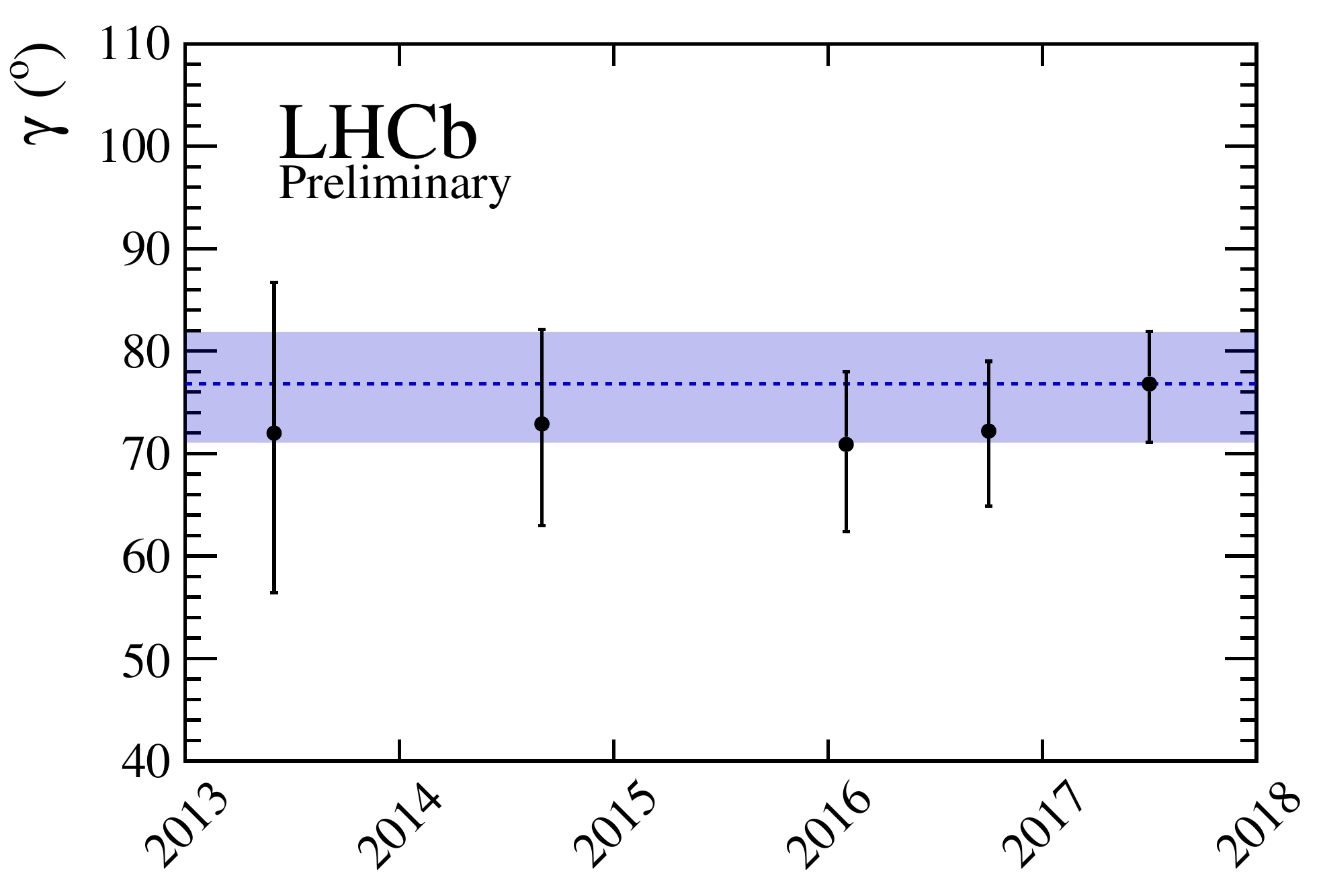}
\caption{\label{Fig:gamma}The evolution of the LHCb $\gamma$ combination.}
\end{figure}

The most precise determination of $\gamma$ ultimately arises through a combination of different analyses
of the suite of different decay modes that have interfering $V_{ub}$ and $V_{cb}$ amplitudes.
In 2016, LHCb obtained the result $\gamma = 72.2^{+6.8}_{-7.3}{\degree}$~\cite{Aaij:2016kjh}, 
through a combination of 89 observables, from analyses of the Run-I dataset.
At the 2017 EPS conference, LHCb presented an updated combination with an improved precision
$\gamma = 76.8^{+5.1}_{-5.7}{\degree}$~\cite{LHCb-CONF-2017-004}.
This includes the $B^- \to D^{(*)0}h^-$ analysis~\cite{Aaij:2017ryw} described above, 
a preliminary update of the $B_s^0 \to D_s^- K^+$ mode~\cite{LHCb-CONF-2016-015}, and a preliminary 
analysis of $B^- \to D^0K^{*-}$ decays~\cite{LHCb-CONF-2016-014}.
Fig.~\ref{Fig:gamma} shows how the precision on the LHCb $\gamma$ combination has improved over the last four
years, up to the summer 2017 version.

\section{Conclusions}

The study of tree-level $b$ decays plays a crucial role in exposing BSM physics effects in the quark flavour sector.
Semileptonic decays provide necessary constraints on the CKM elements $|V_{ub}|$ and $|V_{cb}|$, and the first precise
studies with $b$ baryon decays by LHCb are presented, including a new study of the shape of the differential decay 
rate of the decay $\Lambda_b^0 \to \Lambda_c^+ \mu^- \overline{\nu}$.
Tree-level tests of lepton universality through the $\mathcal{R}(D^{(\ast)})$ observables are discussed,
including the first measurement of $\mathcal{R}(D^{\ast})$ with hadronic three-prong $\tau$ lepton decays, by LHCb.
A LHCb study of $B^- \to D^0h^-$ decays is presented, including a first analysis of $B^- \to D^{\ast0}h^-$ decays 
with a partial reconstruction approach.
Finally, a summer 2017 update of the LHCb $\gamma$ combination is presented.

\bibliographystyle{JHEP}
\bibliography{main}

\end{document}